\documentclass{aa}
\usepackage{graphicx}
\usepackage{latexsym}
\usepackage{amsmath}
\usepackage{amssymb}

\newcommand{\MC}{\multicolumn}
\newcommand{\kms}{km\,s$^{-1}$}
\newcounter{qub}
\newcommand{\qq}{\addtocounter{qub}{1}\arabic{qub}}

\begin{document}

\title{The Hamburg/SAO survey for emission--line galaxies }
\subtitle{VI. The sixth list of 126 galaxies }

\author{%
S.A.~Pustilnik\inst{1}
\and D.~Engels\inst{2}
\and V.A.~Lipovetsky\inst{1}\fnmsep\thanks{Deceased 1996 September 22.}
\and A.Y.~Kniazev\inst{1,3}
\and A.G.~Pramskij\inst{1}
\and A.V.~Ugryumov\inst{1}
\and J.~Masegosa\inst{4}
\and Y.I.~Izotov\inst{5}     
\and F.~Chaffee\inst{6}
\and I.~M\'arquez\inst{4}
\and A.L.~Teplyakova\inst{7}
\and \\ U.~Hopp\inst{8}        
\and N.~Brosch\inst{9}
\and H.-J.~Hagen\inst{2}
\and J.-M.~Martin\inst{10}
}

\offprints{S.~Pustilnik  \email{sap@sao.ru}}

\institute{
Special Astrophysical Observatory, Nizhnij Arkhyz, Karachai-Circassia,
369167, Russia
\and Hamburger Sternwarte, Gojenbergsweg 112, D-21029 Hamburg, Germany
\and European Southern Observatory, Karl-Schwarzschild-Strasse 2, 85748 Garching, Germany
\and Instituto de Astrofisica de Andalucia, CSIC, Aptdo. 3004, 18080, Granada, Spain
\and Main Astronomical Observatory, 27 Zabolotnoho str., Kyiv, 03680, Ukraine
\and Keck Observatories, Hawaii, USA
\and Sternberg Astronomical Institute, Moscow State University, Moscow
\and Universit\"atssternwarte M\"unchen, Scheiner Str. 1, D-81679 M\"unchen, Germany
\and Wise Observatory, Tel-Aviv University, Tel-Aviv 69978, Israel
\and GEPI, Observatoire de Paris, F-92195 Meudon Cedex, France
}

\date{Received 1 December 2004; Accepted 15 June  2005}

\abstract{
We present the sixth list with results\thanks{Tables 4 to 8 are
only available in electronic form at the CDS via anonymous ftp to
cdsarc.u-strasbg.fr (130.79.128.5) or via
http://cdsweb.u-strasbg.fr/Abstract.html. Figures A1 to A13 will be
made available only in the electronic version of the journal.} of the
Hamburg/SAO Survey for Emission-Line Galaxies.
The final list resulted from  follow-up spectroscopy conducted
with the 4.5\,m MMT telescope in 1996, and with  2.2\,m CAHA and 6\,m SAO
telescopes in 2000 to 2003.
The data of this snap-shot spectroscopy survey confirmed
134 emission-line objects out of 182
observed candidates and allowed their
quantitative spectral classification and redshift determination.
We classify 73 emission-line objects as definite or probable blue
compact or H{\sc ii} galaxies (BCG), 8 as QSOs, 4 as Seyfert 1
and 2 galaxies. 30 low-excitation objects were classified as definite
or probable starburst nuclei (SBN), 3 as dwarf amorphous nuclei
starburst galaxies (DANS) and 2 as LINERs.
Due to the low signal-to-noise ratio we could not classify 14 ELGs (NON).
For another 9 galaxies we did not detect any significant
emission lines. For 98 emission-line galaxies, the redshifts and/or
line intensities are determined for the first time.
For the remaining 28 previously-known ELGs we give either improved data
the line intensities or some independent measurements.
The detection rate of ELGs is $\sim$70\%.
This paper completes the classification of strong-lined ELGs
found in the zone of the Hamburg/SAO survey. Together with previously known
BCG/\ion{H}{ii} galaxies in this zone, this sample of $\sim$500
objects is the largest to date in a well bound region.
}

\authorrunning{S.Pustilnik et al.}

\titlerunning{HSS for emission-line galaxies. VI. Sixth list of 126 ELGs}

\maketitle


\begin{table*}
\begin{center}
\caption{\label{Tab1} Journal of observations}
\begin{tabular}{ccccccc} \\ \hline
\MC{1}{c}{ Date } &
\MC{1}{c}{ Telescope }  &
\MC{1}{c}{ Instrument } &
\MC{1}{c}{ Grating,  }  &
\MC{1}{c}{ Wavelength } &
\MC{1}{c}{ Dispersion } &
\MC{1}{c}{ Observed } \\

\MC{1}{c}{ } & & &
\MC{1}{c}{ grism } &
\MC{1}{c}{ range [\AA] } &
\MC{1}{c}{ [\AA/pixel] } &
\MC{1}{c}{ number } \\

\MC{1}{c}{ (1) } &
\MC{1}{c}{ (2) } &
\MC{1}{c}{ (3) } &
\MC{1}{c}{ (4) } &
\MC{1}{c}{ (5) } &
\MC{1}{c}{ (6) } &
\MC{1}{c}{ (7) } \\
\hline
\\[-0.3cm]
20.05.1996       &          4.5~m MMT  & MMT Spect & R300  & 3700--7400 & 3.2 & 42    \\
11-12.04.2000    &          6~m   BTA  & LSS     & R325    & 3600--7600 & 4.6 & 24     \\ 
25.05.2000       &          6~m   BTA  & LSS     & R325    & 3600--7600 & 4.6 &  1     \\ 
28.06-04.07.2000 &          2.2~m CAHA & CAFOS   & G-200   & 3700--9500 & 4.5 & 54     \\ 
03.10-31.10.2000 &          6~m   BTA  & LSS     & R651    & 3700--6000 & 2.3 &  5     \\ 
17.01-20.01.2001 &          6~m   BTA  & LSS     & R651    & 3700--6000 & 2.3 & 16     \\ 
16-19.02.2002    &          6~m   BTA  & LSS     & R400    & 3700--7600 & 3.8 & 15     \\ 
14-15.02.2002    &          2.2~m CAHA & CAFOS   & G-200   & 3700--9500 & 4.5 & 14     \\ 
10.12-13.12.2002 &          6~m   BTA  & LSS     & R400    & 3700--7600 & 3.8 & 10     \\ 
24.12.2003       &          6~m   BTA  & LSS     & R400    & 3700--7600 & 3.8 &  4     \\ %
\hline \\[-0.2cm]
\end{tabular}
\end{center}
\end{table*}


\begin{table*}
\begin{center}
\caption[]{\label{summary} 
Summary of the samples observed and breakdown of the classifications after
follow-up spectroscopy
}
\begin{tabular}{llccccccc}
\hline\noalign{\smallskip}
\MC{2}{c}{Candidate Sample}         &  N  & BCG  & Other & QSO &  Galaxies   & Stars & Not        \\
		    &               &     & \&   & ELGs  &     &  without    &       & Classified \\
		    &               &     & BCG? &       &     &  emission   &       &            \\
\hline\noalign{\smallskip}
First priority      & new           & 36  & 21   &  8    &  1  &  1          &  4    &  1         \\
		    & already known & 40  & 31   &  9    & --  & --          & --    & --         \\
		    & total         & 76  & 52   & 17    &  1  &  1          &  4    &  1         \\[0.10cm]
\hline
Second priority     & new           &  77 & 11   & 17    & 6   & 8           & 22    & 12        \\             
		    & already known &  29 &  9   & 20    & 1   & --          & --    & --         \\[0.10cm]
		    & total         & 106 & 20   & 37    & 7   & 8           & 22    & 12         \\[0.10cm]
\noalign{\smallskip}\hline\noalign{\smallskip}
\MC{2}{l}{Objects presented in this paper}
				    & 182 & 72   & 54    & 8   &  9          & 26    & 13         \\
\noalign{\smallskip}\hline
\end{tabular}
\end{center}
\end{table*}

\section{Introduction}

The problem of creating large, homogeneous and deep samples of actively
star-forming low-mass galaxies is very important for several applications
in studies of galaxy evolution and spatial distribution.
Several earlier projects, based on objective prism plates, like the Second
Byurakan Survey (SBS) (Markarian et al. \cite{Markarian83}, Stepanian
\cite{Stepanian94}), the University of Michigan (UM) survey (e.g., Salzer et
al. \cite{Salzer89}) and the Case survey (Pesch et al. \cite{Pesch95},
Salzer et al. \cite{Salzer95}, Ugryumov et al. \cite{Ugryumov98}), as well
as some others (e.g., Kitt Peak International Spectral Survey -- KISS,
Salzer et al. \cite{KISS}, based on CCD detector registration) identified
several thousand emission-line galaxies.
The Hamburg/SAO survey (HSS)  creates a new very large
homogeneous sample of such galaxies in the region of the Northern sky with an
area of some 1700 square degrees. It was initiated, in particular, in order
to close the gap between the sky regions of the SBS and the original Case
survey, and as a result to get the combined sample of low-mass emission-line
galaxies in a very large section of sky suitable for the study of their
spatial distribution.

The basic outline of the HSS and its first results
are described in Paper I (Ugryumov et al. \cite{Ugryumov99}), while
the additional results from the follow-up spectroscopy are given
in papers II, III, IV and V (Pustilnik et al. \cite{Pustilnik99},
Hopp et al. \cite{Hopp00}, Kniazev et al. \cite{Kniazev01}, Ugryumov et al
\cite{Ugryumov01}).
In this, the last paper, we present the results of the follow-up spectroscopy
of another 182
objects selected on the Hamburg Quasar Survey (HQS) prism spectral plates as
ELG candidates. In Table \ref{summary} we show the breakdown of these
objects in the samples of the 1st and 2nd priority group, and the categories
of detected objects as described below. Out of 134 emission-line objects
(galaxies and QSOs) 69 were known as NED objects. For 28 of these galaxies
either only redshift, or also some information on emission lines was
known, mainly from the previous HSS papers. We included such objects in the
presented list since we provide either significantly improved data
or some independent measurements.

The article is organized as follows. In section \ref{Obs_red} we give the
details of the spectroscopic observations and of the data reduction.
In section \ref{Res_follow} the results of the observations are presented in
several tables. In section \ref{Discussion} we
briefly discuss the new data and summarize the current state of the
Hamburg/SAO survey. Throughout this paper a Hubble constant H$_0$ = 75
km$\,$s$^{-1}$ Mpc$^{-1}$ is used.

\section{Spectral observations and data reduction}
\label{Obs_red}

\subsection{Observations}

The results presented here were obtained mostly in snap-shot observing
mode during one run with the 4.5\,m Multiple Mirror Telescope (MMT), two runs
with the Calar Alto 2.2\,m and seven runs with the SAO 6\,m (BTA) telescopes
(see Table \ref{Tab1}).

\subsection{Observations with the MMT 4.5\,m telescope}

The observations were carried out on May 20, 1996, with the
Red Channel of the MMT Spectrograph
through the long slit of 1\farcs5$\times$180$''$.
The 300 grooves~mm$^{-1}$ grating in first order provides {\bf a} dispersion
of 3.2~\AA~pixel$^{-1}$, and a spectral resolution FWHM of about 10~\AA.
To avoid
second-order contamination, a L-38 blocking filter was used.
The total spectral range was $\lambda\lambda$3700--7400~\AA.
The spectra were rebinned by a factor of 2 along the spatial axis. Hence, the
spatial sampling was 0\farcs6~pixel$^{-1}$.

Short exposures (3--5 minutes) were taken in order to detect strong
emission
lines to allow redshift measurements  and a crude classification.
The slit was not oriented along the parallactic angle  because
of the snap-shot observing mode.
Reference spectra of an Ar--Ne--He lamp were recorded to provide
wavelength calibration.
Spectrophotometric standard stars from Oke (\cite{Oke90}) and Bohlin
(\cite{Bohlin96}) were observed at the beginning and at the end of the
night
for flux calibration. The dome flats, bias, dark and twilight sky frames
were accumulated each night.
The weather conditions were photometric, with seeing variations around
1\farcs0 (FWHM).

\subsection{Calar Alto 2.2\,m telescope observations}
\label{CAFOS}

Follow-up spectroscopy with the CAHA 2.2\,m telescope was carried out
during two runs (June-July 2000 and February 2002), using
the Calar Alto Faint Object Spectrograph (CAFOS).
During these runs a long slit of $300\arcsec \times 2\arcsec $
and a G-200 grism (187\,\AA\,mm$^{-1}$, first order) were used.
The spatial scale along the slit was $0\farcs53$~pixel$^{-1}$.
A SITE~15 2K$\times$2K CCD was operated without binning.
The wavelength coverage was $\lambda$\,3700 -- $\lambda$\,9500\,\AA\
with maximum sensitivity at $\sim$~6000\,\AA.
The spectral resolution was $\sim$~12--16\,\AA\ (FWHM).
The slit orientation was not aligned with the parallactic angle because of
the snap-shot observing mode. The exposure times varied within $5-20$ minutes
depending on the object brightness and weather conditions. The observations
were complemented
by standard star flux measurements (Oke \cite{Oke90}, Bohlin \cite{Bohlin96}),
reference spectra (Hg--Cd lamp) for wavelength calibration, dome flat,
bias and dark frames.
In the run of June-July 2000 the weather conditions were photometric most
of the time with a seeing $\approx$1.5\arcsec\ (FWHM). During one night of
this run,
as well as during two nights in February 2002, the weather conditions were
variable with a seeing of 3\arcsec\ -- 4\arcsec. The measurements in these
nights are marked by ``$*$'' in Table \ref{Tab4}.

There was no order separation filter applied,  therefore
 some second order contamination by the object UV light might be present
at wavelengths longer than 7200~\AA. However, as
can be directly seen from the presented spectra, this effect is probably
small, since it is undetectable in the continuum behavior around
$\lambda$7200~\AA.
In principle, one could expect an increase of the line
fluxes at wavelengths longer than this
due to the second order contamination in the spectra of the flux
calibrating stars.
There are seven objects whose emission line ratios could be potentially
affected. These objects are listed and commented
at the end of section \ref{ELGs}.

\subsection{BTA 6\,m telescope observations}

The observations with the 6\,m telescope (BTA) of the Special Astrophysical
Observatory of Russian Academy of Sciences (SAO RAS) were performed mainly
as a back-up program. Therefore the weather conditions in most cases were
rather poor. The seeing in the majority of the nights was in the
range of
2\arcsec\ to 4\arcsec\ (FWHM) and/or the transparency was variable.
Results obtained under non-photometric conditions are marked by
``$*$'' in Table \ref{Tab4}.
In all cases we used the long slit spectrograph (LSS) in the BTA prime
focus (Afanasiev et al. \cite{Afanasiev95}) with a Photometrics
1K$\times$1K CCD detector with 24~$\mu$m pixel size. The long slit
of 120\arcsec\ was used with the slit width of either 1\farcs5 or 2\farcs0,
depending on the seeing and grating. Three set-ups with the
gratings of 325, 400 and 651  grooves mm$^{-1}$ were used during various runs.
The wavelength ranges of the spectra covered for different set-ups and their
samplings in \AA~pixel$^{-1}$ are given in Table~\ref{Tab1}.
The respective effective resolutions were $\sim$14~\AA, $\sim$11~\AA\ and
$\sim$7~\AA.

Reference spectra of an Ar--Ne--He lamp were recorded before or after
each observation to provide  wavelength calibration.
Spectrophotometric standard stars from Bohlin (\cite{Bohlin96}) were
observed for flux calibration.  All observations were
conducted mainly with the software package {\tt NICE} in MIDAS,
described by Kniazev \& Shergin (\cite{Kniazev95}).

\subsection{Data reduction}

The reduction of all data was performed at SAO
using the standard reduction systems MIDAS\footnote{ MIDAS is an acronym
for the European Southern Observatory package -- Munich Image Data Analysis
System.} and IRAF\footnote[2]{IRAF is distributed by
National Optical Astronomical Observatory, which is operated by the
Association of Universities for Research in Astronomy, Inc., under
cooperative agreement with the National Science Foundation}.

The MIDAS command FILTER/COSMIC was found to be a quite successful
way to automatically remove all cosmic ray hits from the images.
After that we applied the IRAF package CCDRED for bad pixel removal,
trimming, bias-dark subtraction, slit profile and flat-field corrections.

To do accurate wavelength calibration, correction for distortion
and tilt for each frame, sky subtraction and correction for
atmospheric extinction, the IRAF package LONGSLIT was used.

To obtain an instrumental response function from observed
spectrophotometric flux standards,
we first extracted the apertures of standard stars.
Then the determined sensitivity curve was applied
to perform flux calibration for all object images. Finally
we extracted one-dimensional spectra from the flux calibrated images.
When more than one exposure was taken with the same setup
for a given object, the extracted spectra were co-added and a mean vector
was calculated.
When several observations with different setups (telescopes or grisms)
for the same object were obtained, the data were reduced and measured
independently and the more accurate values were taken.

To speed-up and facilitate the line measurements we employed the
dedicated command files created at SAO using the FIT context and MIDAS
command language.
The procedures for the measurements of line parameters and redshifts
applied were also described in detail in Papers III, IV and
in Kniazev et al. (\cite{Kniazev04}).

\section{Results of follow--up spectroscopy}
\label{Res_follow}

In Table~\ref{summary} we present the summary of the observation results.
182 candidates were selected from our first and second priority
samples introduced in Paper IV.

Out of 76 first priority candidates (objects showing a clear density peak
near $\lambda$5000~\AA\ and a blue continuum on the HQS prism spectral
plates), 36 objects appeared in our list as
new ones. 40 objects were listed in the NED as galaxies or objects
from various catalogs and 4 of them already had
information on emission lines and redshifts in earlier publications. Apart
from these 4 objects, 24 more of the mentioned 40 NED galaxies
have appeared in our previous HSS papers, but had data of rather low quality.
All such objects were included in our observing program in order to improve
spectral  information.
The comparison of our measured velocities with those of galaxies with
already known
redshift shows acceptable consistency for most objects in common within
the uncertainties given. However, for five galaxies originally appearing in
the HSS List II, the difference found is as high as 200--300~\kms,
which probably indicates the lower accuracy of some radial
velocities from that list.

The remaining 106 observed objects were taken from the list of the second
priority candidates, those with less prominent emission features
on the high resolution spectra (HRS) obtained after scanning the original
HQS objective prism plates.
As described in Paper IV, we created from this list
the ``APM selected sample'', which uses additional information for
the selection.
The ``APM selected'' sample comprises second priority candidates which
are classified as non-stellar (at least in one of two filters) on Palomar
Sky Survey plates (PSS) in the APM
database, and have a blue colour according to the APM colour system
($(B-R)_{\rm APM} <$ 1.0).
Here we give the spectral data for 64 of them, that looked like ELGs
or QSOs.
31 more 2nd priority candidates were classified as stars or galaxies
without emission,
and 12 objects with no emission lines were not classified at
all due to poor S/N ratio spectra.

\subsection{Emission-line galaxies}
\label{ELGs}

The parameters of the 126 observed emission line galaxies are listed in
Table~\ref{Tab3}, containing the following information: \\
 {\it column 1:} Number in the Table. \\
 {\it column 2:} The object's IAU-type name with the prefix HS. \\
 {\it column 3:} Right ascension for equinox B1950. \\
 {\it column 4:} Declination for equinox B1950.
The coordinates were measured on direct plates of the HQS
and are accurate to $\sim$ 2$\arcsec$ (Hagen et al. \cite{Hagen95}). \\
 {\it column 5:} Heliocentric velocity and its r.m.s. uncertainty in
km~s$^{-1}$. \\
 {\it column 6:} Apparent B-magnitude obtained by calibration of the digitized
photoplates with photometric standard stars (Engels et al. \cite{Engels94}),
having an r.m.s. accuracy of $\sim$ $0\fm5$ for objects fainter than
m$_{\rm B}$ = $16\fm0$ (Popescu et al. \cite{Popescu96}).
Since the algorithm to calibrate the objective prism spectra is
optimized for point sources, the brightnesses of extended galaxies are
underestimated. The resulting systematic uncertainties are expected to
be as large as 2 mag (Popescu et al. \cite{Popescu96}). For about 30\%
of our objects, B-magnitudes are unavailable at the moment. We present
for them blue magnitudes obtained from the APM database. They are
marked by a ``*'' before the value in the corresponding
column. According to our estimate they are systematically brighter by
$0\fm92$ than the B-magnitudes obtained by calibration of the
digitized photoplates (r.m.s.  $1\fm02$). \\
 {\it column 7:} Absolute B-magnitude, calculated from the apparent
B-magnitude and the heliocentric velocity.
No correction for galactic extinction is made because all objects are
located at high galactic latitudes and the corrections are significantly
smaller than the uncertainties in the magnitudes. \\
 {\it column 8:} Preliminary spectral classification type according to
the spectral data presented in this article. BCG means a galaxy
possessing a characteristic H{\sc ii}-region spectrum with low enough
luminosity (M$_B \geq -$20$^m$). SBN and DANS are galaxies of lower
excitation with a corresponding position in the line ratio diagnostic
diagrams, as discussed in Paper~I. SBN are the brighter fraction of this
type. Here we follow the notation of Salzer et al. (\cite{Salzer89}).
The non-confident classification is followed by "?".
Three objects (HS~0807+4103, HS~1525+4344, HS~1627+3625)
were recognized as Sy 1 galaxies due to the presence of broad Balmer lines
and broad [Fe{\sc ii}] emission. HS~1644+3934 was recognized as a
Seyfert 2 galaxy.
The typical spectrum of low-ionization nuclear emission-line regions
(LINERs) is identified for 2 galaxies.
14 ELGs are difficult to classify, mainly due to low S/N. They are coded
as NON. \\
 {\it column 9:} One or more alternative names, according to the
information from NED.
References are given to the other sources of the
redshift-spectral information indicating that a galaxy is an ELG.

The spectra of all emission-line galaxies are shown in Appendix~A,
which is available only in the electronic version of the journal.

The results of line flux measurements are given in Table~\ref{Tab4}
which contains the following information: \\
 {\it column 1:} Number in the Table. \\
 {\it column 2:} The object's IAU-type name with the prefix HS.
Asterisks refer to the objects observed during non-photometric
conditions. \\
{\it column 3:} Designation of the telescope with which the spectral data
were obtained. `B' means BTA, `C' - Calar Alto 2.2\,m telescope, and `M'
- MMT.   \\
 {\it column 4:} Observed flux (in
10$^{-16}$\,erg\,s$^{-1}$\,cm$^{-2}$) of the H$\beta$\, line.
The accuracy of this and other parameters varies substantially over
the whole table. We divided the relative errors into four intervals:
$\le$5\%, (5--10)\%, (10--20)\% and (20--50)\%. They are marked by the
respective superscripts $a$, $b$, $c$ and $d$ right of each table entree.
For about 40\% of ELGs the Balmer absorptions from the underlying
stellar population can somewhat affect the measured H$\beta$\ emission flux
and the related flux ratios. These objects are marked with ``$\dagger$''.
For several objects with non-detected H$\beta$ emission line, the
fluxes are given for H$\alpha$ and marked by a ``$\ddagger$''.  \\
 {\it columns 5,6,7:} The observed flux ratios [O{\sc ii}]/H$\beta$,
[O{\sc iii}]/H$\beta$ and H$\alpha$/H$\beta$. \\
 {\it columns 8,9:} The observed flux ratios
[N{\sc ii}]\,$\lambda$\,6583\,\AA/H$\alpha$, and
([S{\sc ii}]\,$\lambda$\,6716\,\AA\ + \,$\lambda$\,6731\,\AA)/H$\alpha$. \\
 {\it columns 10,11,12:} Equivalent widths of the lines
[O{\sc ii}]\,$\lambda$\,3727\,\AA, H$\beta$ and
[O{\sc iii}]\,$\lambda$\,5007\,\AA.
\\

\noindent
Below we give comments on some specific cases:

\noindent
{\it HS~1010+4907} and {\it HS~1009+4906} comprise a compact group
($\sim$50 kpc in extent) with a fainter galaxy without evident emission
lines, namely {\it HS~1010+4906} (see Table 7). \\
\noindent
{\it HS~1353+4706} was classified as an M-star in Paper I
(Ugryumov et al. \cite{Ugryumov99}). However, it was suspected that
a wrong object had been observed about 0\farcm3 away. This object
will be referred to as HS 1353+4706A.
The new observations indeed revealed that HS 1353+4706B is a very strong-lined
BCG with very low metallicity (12+$\log$(O/H) = 7.63$\pm$0.03; see Pustilnik
et al. \cite{BCG_abun}).
The M-dwarf HS 1353+4706A has B1950 coordinates
13 53 25.2 +47 06 46  and its brightness is B$>$18.7.   \\
Seven galaxies observed with the Calar Alto 2.2\,m telescope have H$\alpha$,
[\ion{N}{ii}] or [\ion{S}{ii}] lines at $\lambda >$ 7200~\AA. Their fluxes
can be affected  by the second order contamination as pointed out in Sect.
\ref{CAFOS}. For these galaxies (1231+4349, 1235+4108, 1426+3658, 1437+3724,
1439+3704, 1525+4344 and 1614+4450) the affected parameters in
Table \ref{Tab4} can be either the ratio F(H$\alpha$)/F(H$\beta$), or the
line flux F(H$\alpha$), if H$\beta$ was not detected. For the ratios of
F([\ion{N}{ii}])/F(H$\alpha$) and F([\ion{S}{ii}])/F(H$\alpha$) the effect
should be minor since these lines are close in  wavelength.

\subsection{Quasars}

The main criteria applied to search for BCGs are  blue continuum near
$\lambda$\,4000\,\AA\  and a strong emission line, the expected doublet
[O{\sc iii}]\,$\lambda$\,4959,5007\,\AA, in
the wavelength region between 5000\,\AA\ and the sensitivity break of
the Kodak IIIa-J photoemulsion near 5400\,\AA\ (see Paper~I). For this
reason faint QSOs with Ly$\alpha$\,$\lambda$\,1216\,\AA\
redshifted to $z \sim$~3, or with C{\sc iv}\, $\lambda$\,1549\,\AA\ redshifted
to $z \sim$~1.7, or with Mg{\sc ii}\, $\lambda$\,2798\,\AA\
redshifted to $z \sim$~0.8 could be selected as BCG candidates.
In Papers I--V we reported the discovery of a number of such faint
QSOs. They were missed by the Hamburg Quasar Survey since it
is restricted to bright QSOs (B $\leq 17-17.5$). Here we report
the discovery of eight faint (B $\ge 17.5$) QSOs.
For four of them we identified Ly$\alpha$\,$\lambda$\,1216\,\AA\
redshifted to $z \sim$~3 as the line responsible for its selection. Two
objects (HS~1608+3546 and HS 1714+4202) show a broad emission line tentatively
identified as Mg{\sc ii}\,$\lambda$\,2798\,\AA\ at $z \sim$ 0.83--0.84.
Two more quasars with
$z \sim 1.7$ were selected due to the line C{\sc iv}\,$\lambda$\,1549\,\AA.
Since for HS 1203+3811 only one broad line is seen in a rather poor
S/N ratio
spectrum, its identification as Ly$\alpha$ should be considered as tentative.
The data for all eight quasars are presented in
Table \ref{Tab5}.  Finding charts and plots of their spectra can be found
on the www-site of the Hamburg Quasar Survey
(http://www.hs.uni-hamburg.de/hqs.html).

\subsection{Non-emission-line objects}

In total, for 49 candidates no (trustworthy) emission lines were detected.
We divided them into three categories.

\subsubsection{Absorption-line galaxies}

For nine non-emission line objects the signal-to-noise ratio
of our spectra was sufficient to detect absorption lines, allowing
the determination of redshifts. The data are presented in Table \ref{Tab6}.

\subsubsection{Stellar objects}

To separate the stars among the objects with no detectable emission
lines, we cross-correlated a list of the most common stellar features
with the observed spectra.  In total, 26 objects with definite stellar
spectra and redshifts close to zero were identified.
All of them were crudely classified in categories
from definite A-stars to G-stars, with most of them intermediate
between A and F. The data for these stars are presented in Table \ref{Tab7}.

\subsubsection{Non-classified  objects}

It was not possible to classify 13 objects without emission lines.
Their spectra have too low signal-to-noise ratio to detect trustworthy
absorption features, or the EWs of their emission lines are too small.

\section{Discussion}
\label{Discussion}

\subsection{The sixth list}

As a result we have 182 observed candidates preselected on HQS objective
prism plates, out of which 76 were first priority candidates
and 106 were second priority.
134 objects (73~\% of the total) are found to be either ELGs (126)
or quasars (8). 24 of these ELGs were presented in the previous HSS
papers, and were reobserved in order to improve the data quality.

Seventy two out of 126 ELGs ($\sim$57~\%) were classified based on
the character
of their spectra and their luminosity as H{\sc ii}/BCGs or probable BCGs.

14 ELGs are difficult to classify due to their poor signal-to-noise
spectra. Six  more ELGs were classified as Active Galactic Nuclei (AGN):
4 as Seyfert galaxies and 2 as LINERs.
The remaining 33 ELGs are objects with low excitation:
either starburst nuclei galaxies (SBN and probable SBN) or their lower mass
analogs -- dwarf amorphous nuclear starburst galaxies (DANS or probable DANS).

\subsection{Brief summary of the HSS for ELGs}
\label{brief}

Summarizing the results of the Hamburg/SAO survey presented in Papers I
through VI, we discovered altogether, from the 1-st priority candidates,
463 new emission-line objects (26 of them are QSOs). For 100 more ELGs
known from the literature (NED) we obtained quantitative data for
their emission lines.
The total number of confident or probable blue compact/H{\sc ii}-galaxies
is 387. Relative to all observed 537 ELGs the fraction of BCGs is $\sim$72\%.

   \begin{figure*}
   \centering
   \includegraphics[width=5.9cm,angle=-90,clip=]{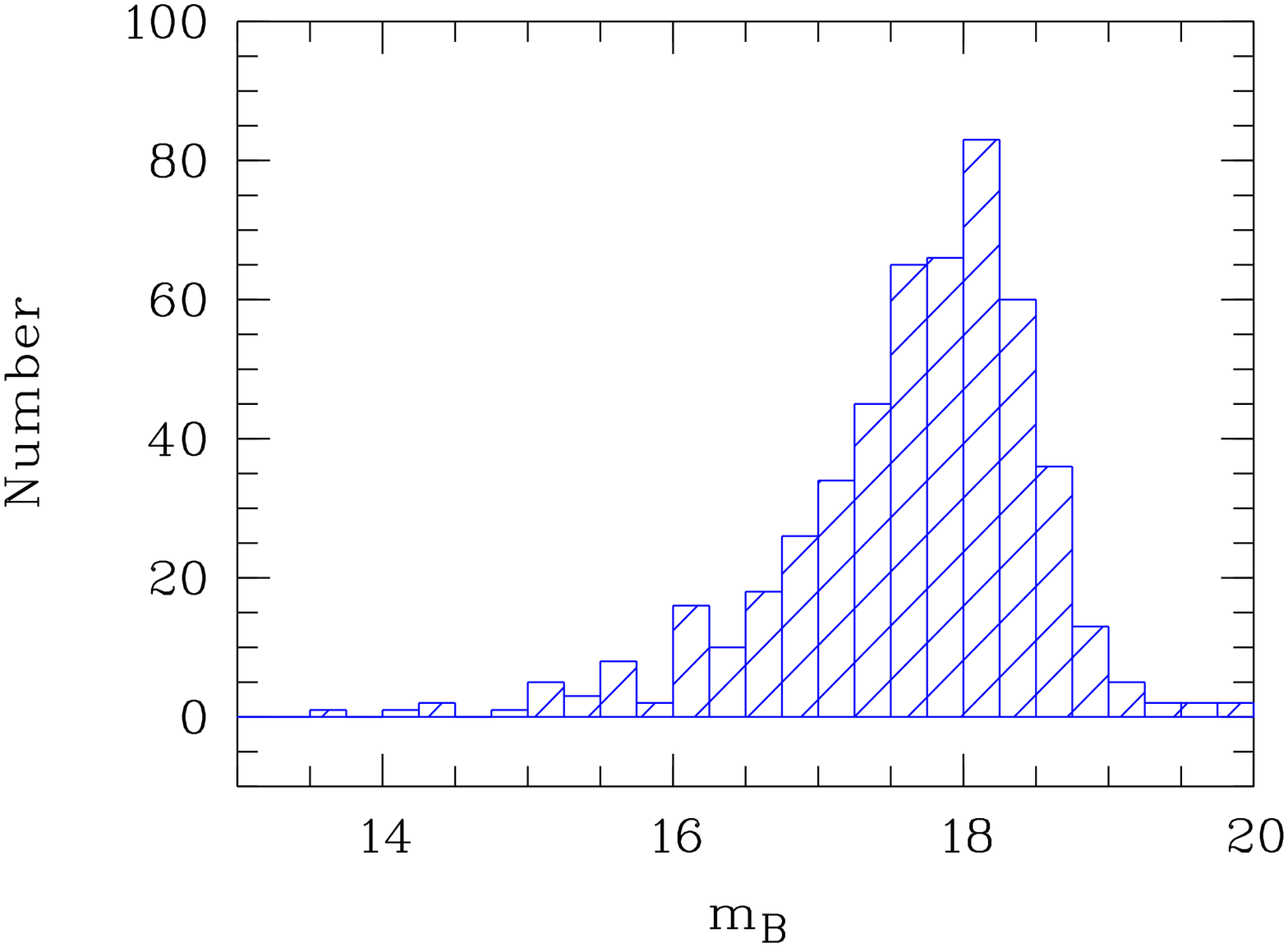}
   \includegraphics[width=5.9cm,angle=-90,clip=]{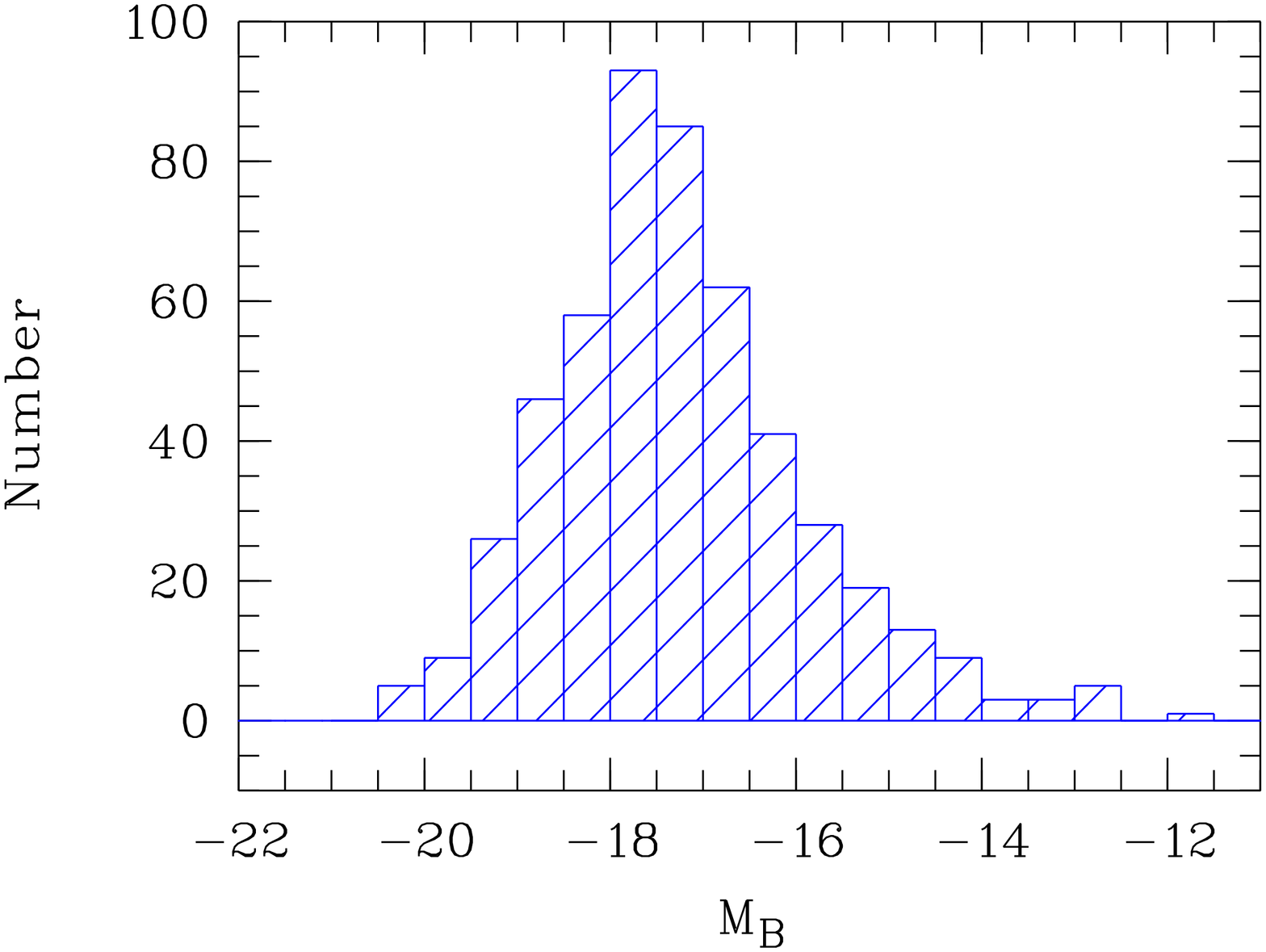}
   \includegraphics[width=5.9cm,angle=-90,clip=]{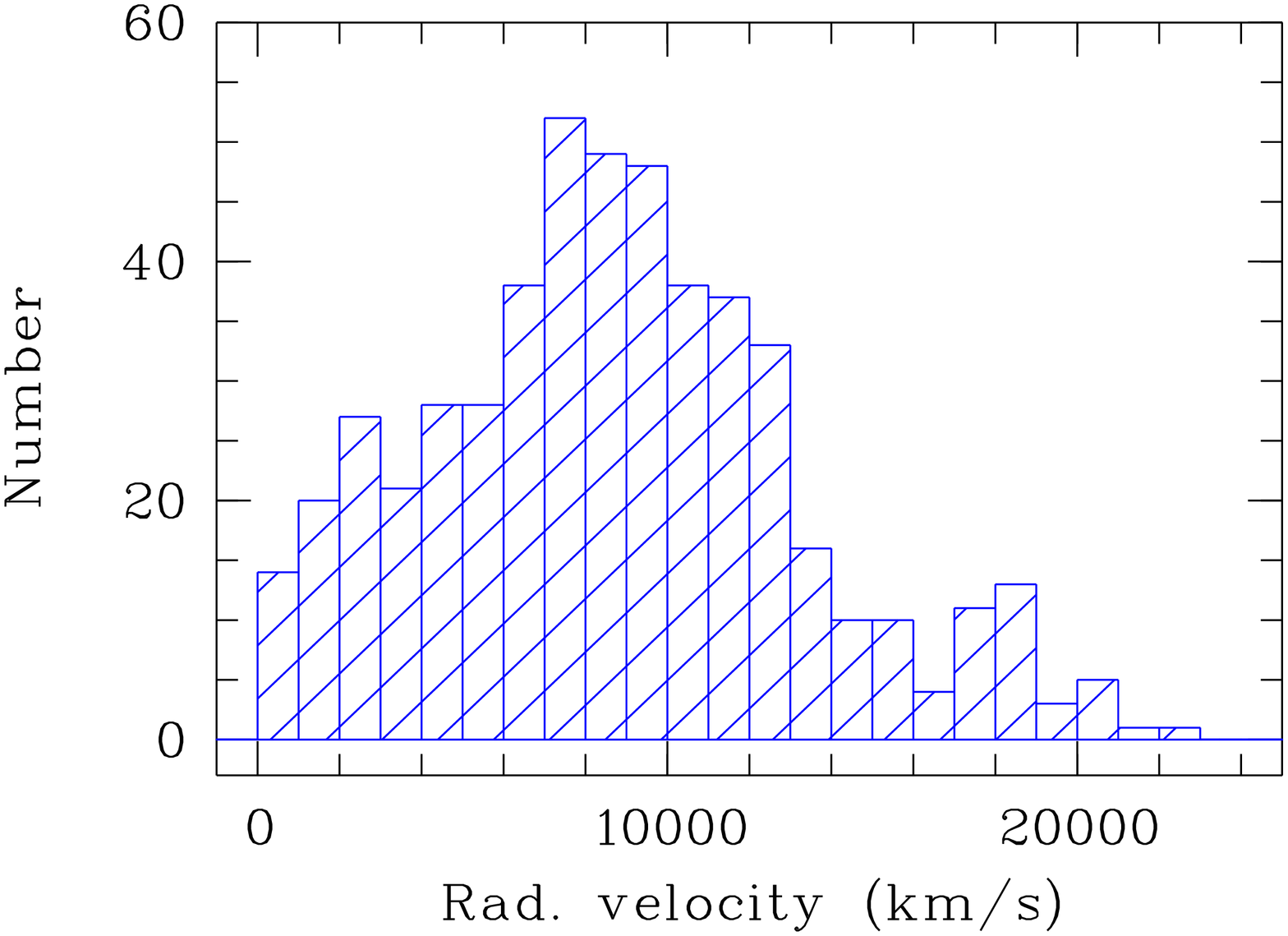}
   \includegraphics[width=5.9cm,angle=-90,clip=]{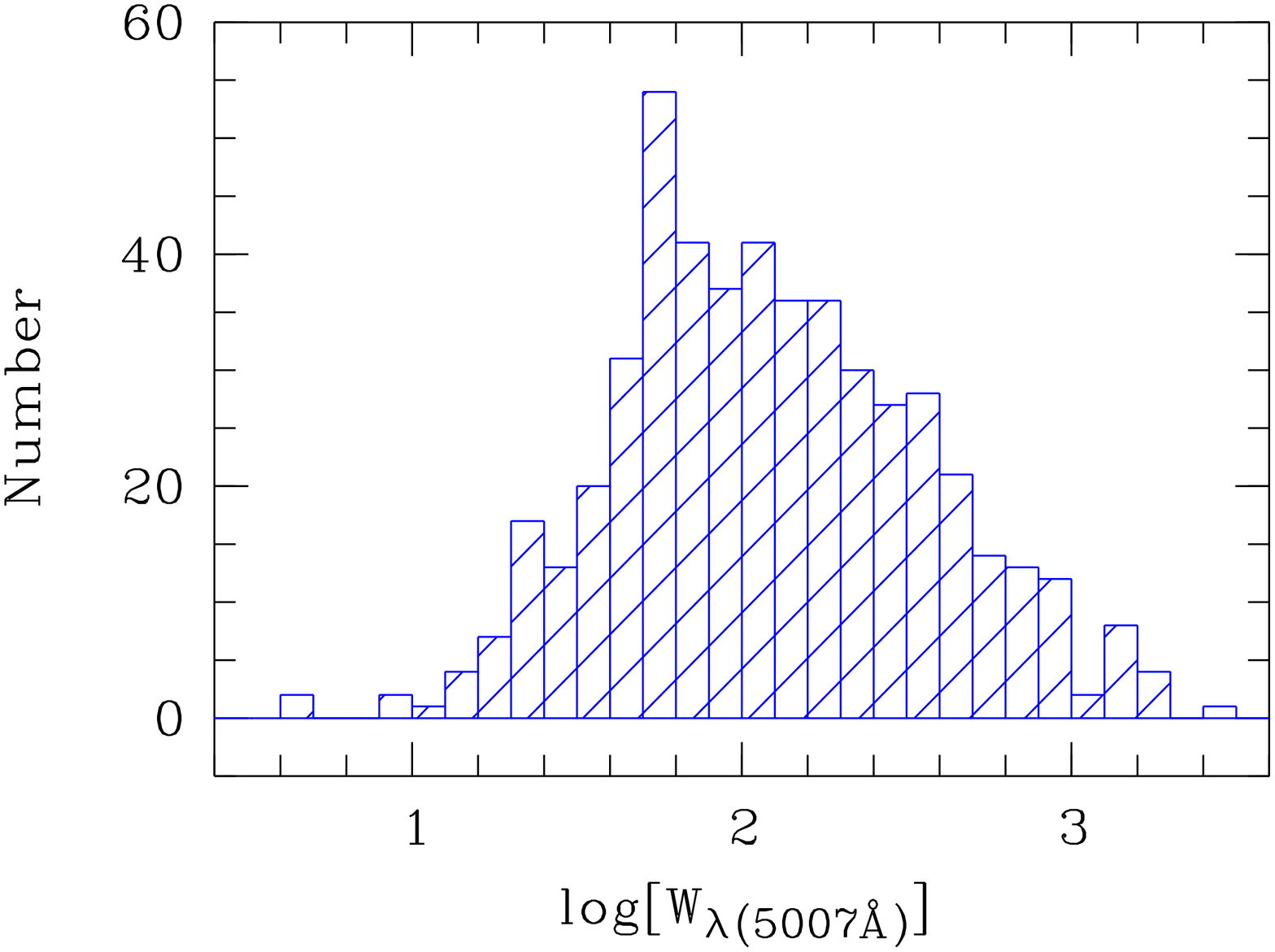}
      \caption{
	 Distributions of the BCG sample in the zone of HSS of
       a) total $B$-magnitude, b) absolute $B$-magnitude, c) radial velocity
	 and  d) equivalent width of [\ion{O}{iii}~5007] line.
	      }
\label{BCG_hist}
   \end{figure*}

42 more new BCGs and 56 other ELGs are found among the second priority
candidates.
Along with the BCGs selected from the HSS candidates, but not observed
by us
since they already were known from other surveys in this region, the total
number of BCGs in the sky region covered by the HSS ($\sim$1700 sq.degrees
of a single piece of sky) reaches $\sim$500.
This constitutes the largest and deepest BCG sample in both hemispheres and
will be presented elsewhere as a separate publication.
The assembly and verifying of the whole HSS database is underway,
and the most up-dated version of this will appear at: http://precise.sao.ru.

In Figure \ref{BCG_hist}(a-d) we show the distributions of radial velocities,
apparent and absolute magnitudes and EWs([\ion{O}{iii}]~5007)
for all found BCGs or BCGs? in the zone of HSS (506 objects, including also
the galaxies, selected by us as the HSS candidates, but not observed
in this project due to lack of observing time;  these
objects were already known/classified and had redshift data). The latter
group comprises about 70 galaxies which came mainly from the papers by
Peimbert \& Torres-Peimbert (\cite{PTP92}), Vogel et al. (\cite{Vogel93}),
Popescu et al. (\cite{Popescu96,Popescu97,Popescu_L2}), Ugryumov et al.
(\cite{Ugryumov98}), Popescu \& Hopp (\cite{Popescu00}) and some unpublished
data on BCGs in the SBS zone, partly intersecting the HSS zone.

Since for most of the sample galaxies the total $B$-band magnitudes are not
available, we used their photographic blue magnitudes from the Palomar Sky
Survey, as provided in the APM database (Automatic Plate-measuring
Machine
at Cambridge, Irwin \cite{Irwin98}), and calibrated them through the sample
of about a hundred BCGs with  CCD-measured total $B$ magnitudes.
We obtained the linear regression of:
\begin{equation}
 B_{\rm CCD} = 0.429 \times B_{\rm APM} + 10.51,
\end{equation}
in the range of $B_{\rm APM}$ from 14 to 19.5,
with the standard deviation of residuals of 0.43 mag.
These $B$ magnitudes were used as a first approximation to more
accurate data
in order to estimate absolute magnitudes and to look at the distributions
of the sample galaxies of these parameters.

\begin{table}
\begin{center}
\caption[]{\label{histo_summary}
Characteristics of the HSS BCG sample
}
\begin{tabular}{llrrr}
\hline\noalign{\smallskip}
 Parameter          &  Number   & Mean  &Median  &80\%~~~~    \\
		    &  of BCGs  &       &        &interval~   \\
\hline\noalign{\smallskip}
B$_{\rm tot}$ (mag) & 506       & 17.66 &17.81   & 16.51,18.59~ \\
M$_{B}$       (mag) & 506       &--17.25&--17.40 &--15.40,--18.89\\
V$_{\rm hel}$ (\kms)& 506       & 8471  &8752    & 2695,14976    \\
EW(5007)(\AA)       & 502       & 122   &113     & 33,524~~    \\[0.10cm]
\noalign{\smallskip}\hline
\end{tabular}
\end{center}
\end{table}

The distributions shown in Figure \ref{BCG_hist} have the characteristics
presented in Table \ref{histo_summary}.
A rough estimate of the completeness level
from the shown distribution is $B_{\rm tot}$ = 18.0.
The radial velocity distribution peaks near V=7500~\kms,
while the distribution of EW(5007) peaks at the value of $\sim$60~\AA.

\subsection{Use of the HSS ELG sample}

Since the new BCG sample is the largest sample of low-mass galaxies
and is well situated on the sky, it can be used for several purposes.
First, as it was assumed in planning this survey, such a sample is suitable
to address the problem of the spatial distribution of low-mass galaxies
relative to the structures delineated by bright galaxies.
A more detailed study  of the already-known differences between luminous and
faint galaxies (e.g., Salzer \cite{S89}; Pustilnik et al. \cite{PULTG};
Popescu et al. \cite{Popescu97}) could help to gain a deeper
understanding of the
CDM structure N-body simulations (e.g., such as by Mathis \& White
\cite{MW02} and  Gottl\"ober et al. \cite{Gott03}).

One of the aims of the HSS project was to close the gap between the
sky regions of the SBS and Case survey. This goal has now been reached.
Since the
HSS has intersections with both, the possible differences in their
selection
functions and other sample characteristics can be quantified and accounted
for. Thus, the
useful ranges of galaxy parameters for which ELGs can be studied in the
whole region of sky covered by the SBS, HSS and Case will be obtained.

Another important aspect of the BCG studies is related to the starburst
triggering mechanisms. It can be addressed with the HSS sample to check the
preliminary conclusion about the important role of galaxy interactions
made, e.g., on the large sample of the SBS BCGs (Pustilnik et al.
\cite{Pustilnik01}). Similar studies have been conducted on the samples
from the
2dF Survey project by Lambas et al. (\cite{Lambas03}) and Alonso et al.
(\cite{Alonso04}).

One more interesting aspect of statistical studies of this BCG sample is
related to the high S/N ratio spectra of the subsample of the most
strong-lined ELGs. This allows us to determine in a large sample of galaxies
the abundance of oxygen and other heavy elements by the classic $T_{\rm e}$
method, and to use these data to compare the BCG properties with the models
of galaxy chemical evolution. We already obtained such data for a
significant fraction of the HSS BCG sample ($\sim$15\% of all BCG and
$\sim$40~\%  BCGs with the EW([\ion{O}{iii}]$\lambda$5007) $> $ 150~\AA,
Pustilnik et al. \cite{BCG_abun}).
Some of the strong-lined HSS BCGs were used in the new
primordial helium determination (Izotov \& Thuan \cite{IzTh04}).

One of the goals of the HSS project was to search for new extremely
metal-deficient (XMD) BCGs (those with 12+$\log$(O/H) $\leq$ 7.65),
possible analogs of candidate young galaxies, like I Zw 18 and SBS 0335--052.
Altogether, in addition to the two XMD galaxies in this zone known from
previous studies (1415+437=CG~389 and 1224+3756=CG~1024),
eight new such galaxies are found (see papers by
Kniazev et al. \cite{KPU98, Kniazev_00, Kniazev_HS0822}, Pustilnik et al.\,
\cite{HS0837, BCG_abun}, Guseva et al.\, \cite{Guseva03}).
Thus, the fraction of XMD BCGs at the magnitude
limit of the HSS is $\sim$2\% (as already claimed by Pustilnik et al.
\cite{Kiel02}), about $\sim$1.5 times higher than the fraction found
by Kniazev et al. (\cite{Kniazev03}) for the Sloan Digital Sky Survey (SDSS).
While we are dealing with small samples in
either case, which makes this ratio uncertain, the ratio still
indicates that we succeeded in creating a design for the HSS which is
more sensitive in
finding XMD BCGs than general galaxy surveys like the SDSS.

\section{Conclusions}

We performed the follow-up spectroscopy of the sixth and last list of
candidates for ELGs (mainly of \ion{H}{ii} type) from the  Hamburg/SAO Survey.
Summarizing the results, the analysis of the spectral
information and the discussion above we draw the following conclusions:

\begin{itemize}

\item The methods to detect ELG candidates on the plates of the
      Hamburg Quasar Survey give a reasonably high detection rate of
      \ion{H}{ii} type emission-line objects. In total,  within the
      two defined  priority categories,
      182 objects were observed, 27 of which were already known as ELGs.
      Among the remaining 155 objects we found 107 emission-line
      objects corresponding to a detection rate of $\sim$~68~\%.

\item Besides ELGs we also found 8 new quasars, with either
      Ly$\alpha$, or C{\sc iv}$\lambda$1549, or Mg{\sc ii}$\lambda$2798
      in the wavelength region $4950-5100$\,\AA\
      near the red boundary of the IIIa-J photoplates
      (z~$\sim$~3, 1.7 and 0.8, respectively).

\item The fraction of BCG/H{\sc ii} galaxies among all new observed
      ELGs (about 43~\%) is lower in this paper compared to the
      previous parts
      of the HSS since about 2/3 of the observed candidates came from
      the second priority list.

\item This list completes the classification work on the strong-lined ELGs
      in the zone of the Hamburg/SAO survey. Together with previously-known
      BCG/\ion{H}{ii} galaxies in this zone, this sample of $\sim$500
      objects is the largest one made to date in a well bound region.
\end{itemize}

\begin{acknowledgements}

This work was supported by the grant of the Deutsche
Forschungsgemeinschaft No.~436 RUS~17/77/94 and by the Russian Federal
Program "Astronomy". U.A.V. is very grateful
to the staff of the Hamburg Observatory for their hospitality and kind
assistance.  Support by the INTAS grant No.~96-0500 is gratefully
acknowledged. I.M. and J.M. acknowledge financial
support by DGICyT grants AYA2001-2089 and AYA2003-00128 and the Junta de 
Andaluc\'{\i}a. The authors thank the anonymous referee for useful
comments and suggestions.
The use of APM facility was very important
for selection methods for additional candidates to BCGs from the 2nd
priority list.
This research has made use of the
NASA/IPAC Extragalactic Database (NED) which is operated by the Jet
Propulsion Laboratory, California Institute of Technology, under contract
with the National Aeronautics and Space Administration. We have
also used the Digitized Sky Survey, produced at the Space Telescope Science
Institute under government grant NAG W-2166.

\end{acknowledgements}

\clearpage

\renewcommand{\baselinestretch}{1.3}


\scriptsize
\setcounter{qub}{0}

\begin{table*}[h]

\begin{center}
\caption{\label{Tab3} Coordinates, velocities and magnitudes of
emission--line galaxies}

\begin{tabular}{rlllrrcll} \hline \\[-0.35cm]
\multicolumn{1}{c}{\#}               &
\multicolumn{1}{c}{Name}             &
\multicolumn{1}{c}{$\alpha\,(1950)$} &
\multicolumn{1}{c}{$\delta\,(1950)$} &
\multicolumn{1}{c}{$V_{hel}^{\,a}\pm\sigma_V$}    &
\multicolumn{1}{r}{m$_{B}$}          &
\multicolumn{1}{c}{M$_{B}^{\;\;b}$}  &
\multicolumn{1}{c}{Type}             &
\multicolumn{1}{l}{Other names from NED}\\
&&&&&&&& \multicolumn{1}{l}{and number of reference}\\
&
\multicolumn{1}{c}{ (2) } &
\multicolumn{1}{c}{ (3) } &
\multicolumn{1}{c}{ (4) } &
\multicolumn{1}{c}{ (5) } &
\multicolumn{1}{c}{ (6) } &
\multicolumn{1}{c}{ (7) } &
\multicolumn{1}{c}{ (8) } &
\multicolumn{1}{c}{ (9) } \\
\hline
\qq& HS 0713+4105$^\ddagger$  &07 13 28.5 &+41 05 51  &  13768 $\pm$  40 &$^\star$17.00 &--19.31 &BCG    & 2MASX J07165730+4100327   \\ 
\qq& HS 0719+4928$^\ddagger$  &07 19 00.7 &+49 28 19  &  18659 $\pm$  40 &        17.60 &--19.37 &SBN?   &                           \\ 
\qq& HS 0727+4711$^\ddagger$  &07 27 52.1 &+47 11 51  &  20192 $\pm$  30 &        18.00 &--19.15 &SBN?   & 2MASX J07313250+4705296   \\ 
\qq& HS 0732+3529  &07 32 23.3 &+35 29 07  &   3785 $\pm$  42 &        17.80 &--15.71 &BCG?   & {\bf (1)}                   \\
\hline
\end{tabular}
\end{center}
\end{table*}

\newpage

\begin{table*}[h]

\begin{center}
\flushleft {\bf Table 4.} (Continued)

\begin{tabular}{rlllrrcll} \\[-0.22cm] \hline \\[-0.35cm]
\multicolumn{1}{c}{\#}               &
\multicolumn{1}{c}{Name}             &
\multicolumn{1}{c}{$\alpha\,(1950)$} &
\multicolumn{1}{c}{$\delta\,(1950)$} &
\multicolumn{1}{c}{$V_{hel}^{\,a}\pm\sigma_V$}    &
\multicolumn{1}{r}{m$_{B}$}          &
\multicolumn{1}{c}{M$_{B}^{\;\;b}$}  &
\multicolumn{1}{c}{Type}             &
\multicolumn{1}{l}{Other names from NED}\\
&&&&&&&& \multicolumn{1}{l}{and number of reference}\\
&
\multicolumn{1}{c}{ (2) } &
\multicolumn{1}{c}{ (3) } &
\multicolumn{1}{c}{ (4) } &
\multicolumn{1}{c}{ (5) } &
\multicolumn{1}{c}{ (6) } &
\multicolumn{1}{c}{ (7) } &
\multicolumn{1}{c}{ (8) } &
\multicolumn{1}{c}{ (9) } \\
\hline
\qq& HS 1121+4707  &11 21 02.4 &+47 07 25  &  10490 $\pm$  38 &        19.30 &--16.42 &BCG    &                          \\ 
\qq& HS 1122+4230  &11 22 10.3 &+42 30 56  &  13371 $\pm$  30 &$^\star$16.87 &--19.38 &SBN    & KUG 1122+425, CG1424     \\ 
\qq& HS 1123+4716  &11 23 01.9 &+47 16 29  &  10115 $\pm$  46 &        16.50 &--19.15 &BCG    & MRK 0168        {\bf (2)}  \\
\hline
\end{tabular}
\end{center}
\end{table*}

\newpage

\begin{table*}[h]

\begin{center}
\flushleft {\bf Table 4.} (Continued)

\begin{tabular}{rlllrrcll} \\[-0.22cm] \hline \\[-0.35cm]
\multicolumn{1}{c}{\#}               &
\multicolumn{1}{c}{Name}             &
\multicolumn{1}{c}{$\alpha\,(1950)$} &
\multicolumn{1}{c}{$\delta\,(1950)$} &
\multicolumn{1}{c}{$V_{hel}^{\,a}\pm\sigma_V$}    &
\multicolumn{1}{r}{m$_{B}$}          &
\multicolumn{1}{c}{M$_{B}^{\;\;b}$}  &
\multicolumn{1}{c}{Type}             &
\multicolumn{1}{l}{Other names from NED}\\
&&&&&&&& \multicolumn{1}{l}{and number of reference}\\
&
\multicolumn{1}{c}{ (2) } &
\multicolumn{1}{c}{ (3) } &
\multicolumn{1}{c}{ (4) } &
\multicolumn{1}{c}{ (5) } &
\multicolumn{1}{c}{ (6) } &
\multicolumn{1}{c}{ (7) } &
\multicolumn{1}{c}{ (8) } &
\multicolumn{1}{c}{ (9) } \\
\hline
\qq& HS 1333+3522$^\ddagger$  &13 33 30.5 &+35 22 58  &   6691 $\pm$  40 &        17.40 &--17.35 &BCG    & MAPS-NGP O\_270\_0867227 \\ 
\qq& HS 1340+3806$^\ddagger$  &13 40 18.2 &+38 06 02  &   6676 $\pm$  40 &        18.70 &--16.04 &BCG    & NGP9 F270-0161337        \\ 
\qq& HS 1340+4428  &13 40 45.3 &+44 28 47  &  13233 $\pm$  42 &$^\star$17.60 &--18.63 &BCG    & MAPS-NGP O\_220\_0009424 \\
\hline
\end{tabular}
\end{center}
\end{table*}

\newpage

\begin{table*}[h]

\begin{center}
\flushleft {\bf Table 4.} (Continued)

\begin{tabular}{rlllrrcll} \\[-0.22cm] \hline \\[-0.35cm]
\multicolumn{1}{c}{\#}               &
\multicolumn{1}{c}{Name}             &
\multicolumn{1}{c}{$\alpha\,(1950)$} &
\multicolumn{1}{c}{$\delta\,(1950)$} &
\multicolumn{1}{c}{$V_{hel}^{\,a}\pm\sigma_V$}    &
\multicolumn{1}{r}{m$_{B}$}          &
\multicolumn{1}{c}{M$_{B}^{\;\;b}$}  &
\multicolumn{1}{c}{Type}             &
\multicolumn{1}{l}{Other names from NED}\\
&&&&&&&& \multicolumn{1}{l}{and number of reference}\\
&
\multicolumn{1}{c}{ (2) } &
\multicolumn{1}{c}{ (3) } &
\multicolumn{1}{c}{ (4) } &
\multicolumn{1}{c}{ (5) } &
\multicolumn{1}{c}{ (6) } &
\multicolumn{1}{c}{ (7) } &
\multicolumn{1}{c}{ (8) } &
\multicolumn{1}{c}{ (9) } \\
\hline
\qq& HS 1715+4827  &17 15 10.3 &+48 27 25  &   8735 $\pm$  62 &        19.80 &--15.53 &BCG    &                          \\ 
\qq& HS 1720+4720  &17 20 55.3 &+47 20 50  &   7964 $\pm$  45 &        17.80 &--17.33 &SBN    &                          \\ 
\qq& HS 1723+4410  &17 23 40.2 &+44 10 07  &  15789 $\pm$  40 &$^\star$17.55 &--19.06 &SBN?   &2MASX J17250981+4407314   \\
\end{tabular}
\end{center}
\end{table*}


\renewcommand{\baselinestretch}{1.4}

\scriptsize
\setcounter{qub}{0}

\begin{table*}[h]

\begin{center}
\caption{\label{Tab4} {\bf Emission line parameters}}

\begin{tabular}{rlrrrrrrrrrr} \hline \\[-0.35cm]
\multicolumn{1}{c}{\#}                             &
\multicolumn{1}{c}{ Name}                          &
\multicolumn{1}{c}{ T}                             &
\multicolumn{1}{c}{ F(H$\beta$) }              &
\multicolumn{1}{c}{ $\frac{F(3727)}{F(H\beta)}$} &
\multicolumn{1}{c}{ $\frac{F(5007)}{F(H\beta)}$} &
\multicolumn{1}{c}{ $\frac{F(H\alpha)}{F(H\beta)}$} &
\multicolumn{1}{c}{ $\frac{F(6583)}{F(H\alpha)}$}&
\multicolumn{1}{c}{ $\frac{F([SII])}{F(H\alpha)}$} &
\multicolumn{1}{c}{ EW$_{3727}$}                   &
\multicolumn{1}{c}{ EW$_{H\beta}$}                 &
\multicolumn{1}{c}{ EW$_{5007}$}                   \\
&
\multicolumn{1}{c}{ (2) } &
\multicolumn{1}{c}{ (3) } &
\multicolumn{1}{c}{ (4) } &
\multicolumn{1}{c}{ (5) } &
\multicolumn{1}{c}{ (6) } &
\multicolumn{1}{c}{ (7) } &
\multicolumn{1}{c}{ (8) } &
\multicolumn{1}{c}{ (9) } &
\multicolumn{1}{c}{ (10) } &
\multicolumn{1}{c}{ (11) } &
\multicolumn{1}{c}{ (12) } \\
\hline


\qq& HS 0713+4105$*$           & C&41$^b$ &1.84$^d$  & 3.15$^b$  &4.90$^b$    &0.15$^b$   &0.23$^b$   &  33$^c$ &  26$^b$ &  82$^a$  \\
\qq& HS 0719+4928$*\dagger$    & B&14$^b$ &2.92$^c$  & 1.77$^c$  &5.45$^b$    &0.25$^b$   &0.29$^b$   &  27$^b$ &   8$^b$ &  13$^b$  \\
\qq& HS 0727+4711$*$           & B&20$^c$ &2.86$^c$  & 1.12$^c$  &5.53$^c$    &0.37$^b$   &0.29$^b$   &  46$^c$ &  11$^c$ &  12$^c$  \\
\qq& HS 0732+3529              & B&44$^b$ &1.27$^c$  & 6.17$^c$  &            &           &           &  53$^c$ &  55$^b$ & 331$^a$ \\
\qq& HS 0735+4507$^\ddagger$$*$& B&10$^c$ &          &           &            &0.28$^d$   &0.87$^d$   &         &         &          \\
\end{tabular}
\end{center}
\end{table*}

\newpage

\begin{table*}[h]

\begin{center}
\flushleft {\bf Table 5.} (Continued)
\begin{tabular}{rlrrrrrrrrrr} \hline \\[-0.35cm]
\multicolumn{1}{c}{\#}                             &
\multicolumn{1}{c}{ Name}                          &
\multicolumn{1}{c}{ T}                             &
\multicolumn{1}{c}{ F(H$\beta$) }                  &
\multicolumn{1}{c}{ $\frac{F(3727)}{F(H\beta)}$}   &
\multicolumn{1}{c}{ $\frac{F(5007)}{F(H\beta)}$}   &
\multicolumn{1}{c}{ $\frac{F(H\alpha)}{F(H\beta)}$}&
\multicolumn{1}{c}{ $\frac{F(6583)}{F(H\alpha)}$}  &
\multicolumn{1}{c}{ $\frac{F([SII])}{F(H\alpha)}$} &
\multicolumn{1}{c}{ EW$_{3727}$}                   &
\multicolumn{1}{c}{ EW$_{H\beta}$}                 &
\multicolumn{1}{c}{ EW$_{5007}$}                  \\
&
\multicolumn{1}{c}{ (2) } &
\multicolumn{1}{c}{ (3) } &
\multicolumn{1}{c}{ (4) } &
\multicolumn{1}{c}{ (5) } &
\multicolumn{1}{c}{ (6) } &
\multicolumn{1}{c}{ (7) } &
\multicolumn{1}{c}{ (8) } &
\multicolumn{1}{c}{ (9) } &
\multicolumn{1}{c}{ (10) } &
\multicolumn{1}{c}{ (11) } &
\multicolumn{1}{c}{ (12) } \\
\hline

\qq& HS 1121+4707             & M&27$^b$    &          & 3.76$^b$  &3.36$^b$    & $<$0.09   &0.17$^d$   &         &  41$^b$ & 163$^a$  \\ 
\qq& HS 1122+4230             & M&27$^c$    &2.48$^c$  &1.04$^c$   &6.87$^c$    &0.32$^b$   &0.36$^c$   &  26$^c$ &   9$^c$ &   9$^c$  \\ 
\qq& HS 1123+4716$\dagger$    & B&152$^a$   &1.49$^a$  & 3.48$^a$  &            &           &           &  97$^a$ &  50$^a$ & 160$^a$  \\
\qq& HS 1128+4753$\dagger$    & C&12$^b$    &          & 4.88$^b$  &3.07$^b$    & $<$0.11   &0.13$^c$   &         &  15$^b$ &  82$^a$  \\
\qq& HS 1134+3639$*\dagger$   & B&34$^a$    &          & 3.78$^a$  &            &           &           &  98$^d$ &  18$^a$ &  66$^a$  \\
\end{tabular}
\end{center}
\end{table*}

\newpage

\begin{table*}[h]

\begin{center}
\flushleft {\bf Table 5.} (Continued)
\begin{tabular}{rlrrrrrrrrrr} \hline \\[-0.35cm]
\multicolumn{1}{c}{\#}                             &
\multicolumn{1}{c}{ Name}                          &
\multicolumn{1}{c}{ T}                             &
\multicolumn{1}{c}{ F(H$\beta$) }                  &
\multicolumn{1}{c}{ $\frac{F(3727)}{F(H\beta)}$}   &
\multicolumn{1}{c}{ $\frac{F(5007)}{F(H\beta)}$}   &
\multicolumn{1}{c}{ $\frac{F(H\alpha)}{F(H\beta)}$}&
\multicolumn{1}{c}{ $\frac{F(6583)}{F(H\alpha)}$}  &
\multicolumn{1}{c}{ $\frac{F([SII])}{F(H\alpha)}$} &
\multicolumn{1}{c}{ EW$_{3727}$}                   &
\multicolumn{1}{c}{ EW$_{H\beta}$}                 &
\multicolumn{1}{c}{ EW$_{5007}$}                  \\
&
\multicolumn{1}{c}{ (2) } &
\multicolumn{1}{c}{ (3) } &
\multicolumn{1}{c}{ (4) } &
\multicolumn{1}{c}{ (5) } &
\multicolumn{1}{c}{ (6) } &
\multicolumn{1}{c}{ (7) } &
\multicolumn{1}{c}{ (8) } &
\multicolumn{1}{c}{ (9) } &
\multicolumn{1}{c}{ (10) } &
\multicolumn{1}{c}{ (11) } &
\multicolumn{1}{c}{ (12) } \\
\hline

\qq& HS 1333+3522$*\dagger$   & B&10$^d$    &          & 4.72$^d$  &6.52$^d$    &$<$0.12    &0.29$^d$   &         &   5$^d$ &  26$^b$  \\
\qq& HS 1340+3806$*$          & B&11$^d$    &          & 2.45$^d$  &4.29$^d$    &    $<$0.16&0.36$^c$   &         &   8$^d$ &  20$^c$  \\
\qq& HS 1340+4428             & M&61$^a$    &2.21$^b$  & 4.71$^a$  &3.25$^a$    &0.04$^d$   &0.14$^c$   & 124$^b$ &  44$^a$ & 193$^a$  \\ 
\qq& HS 1341+4318             & M&29$^a$    &          & 6.07$^a$  &3.26$^a$    &0.04$^d$   &0.12$^d$   &         & 122$^a$ & 622$^a$  \\ 
\qq& HS 1353+4706             & B&14$^a$    &0.28$^c$  & 6.35$^a$  &2.79$^a$    &    $<$0.03&0.01$^d$   &  55$^c$ & 223$^a$ &1880$^a$  \\
\end{tabular}
\end{center}
\end{table*}

\newpage

\begin{table*}[h]

\begin{center}
\flushleft {\bf Table 5.} (Continued)
\begin{tabular}{rlrrrrrrrrrr} \hline \\[-0.35cm]
\multicolumn{1}{c}{\#}                             &
\multicolumn{1}{c}{ Name}                          &
\multicolumn{1}{c}{ T}                             &
\multicolumn{1}{c}{ F(H$\beta$) }                  &
\multicolumn{1}{c}{ $\frac{F(3727)}{F(H\beta)}$}   &
\multicolumn{1}{c}{ $\frac{F(5007)}{F(H\beta)}$}   &
\multicolumn{1}{c}{ $\frac{F(H\alpha)}{F(H\beta)}$}&
\multicolumn{1}{c}{ $\frac{F(6583)}{F(H\alpha)}$}  &
\multicolumn{1}{c}{ $\frac{F([SII])}{F(H\alpha)}$} &
\multicolumn{1}{c}{ EW$_{3727}$}                   &
\multicolumn{1}{c}{ EW$_{H\beta}$}                 &
\multicolumn{1}{c}{ EW$_{5007}$}                   \\
&
\multicolumn{1}{c}{ (2) } &
\multicolumn{1}{c}{ (3) } &
\multicolumn{1}{c}{ (4) } &
\multicolumn{1}{c}{ (5) } &
\multicolumn{1}{c}{ (6) } &
\multicolumn{1}{c}{ (7) } &
\multicolumn{1}{c}{ (8) } &
\multicolumn{1}{c}{ (9) } &
\multicolumn{1}{c}{ (10) } &
\multicolumn{1}{c}{ (11) } &
\multicolumn{1}{c}{ (12) } \\
\hline

\qq& HS 1715+4827$\dagger$  & M&41$^a$    &2.15$^d$  &  4.48$^a$ & 2.89$^a$   &    $<$0.06& 0.11$^d$  &374$^d$    & 69$^a$    & 235$^a$    \\
\qq& HS 1720+4720           & M&61$^a$    &3.68$^b$  &  1.74$^a$ & 3.98$^a$   & 0.19$^b$  & 0.19$^b$  &227$^b$    & 31$^a$    &  54$^a$    \\
\qq& HS 1723+4410$^\ddagger$& M&16$^c$    &          &           &            &0.27$^d$   & 0.52$^d$  &       &       &        \\
\qq& HS 1724+4001$^\ddagger$& M&22$^a$    &          &           &            &0.32$^d$   & 0.50$^d$  &       &       &        \\
\qq& HS 1730+4601           & M&105$^a$   &0.59$^c$  &  6.52$^a$ & 3.16$^a$   &    $<$0.03& 0.05$^d$  & 39$^c$    &159$^a$    & 915$^a$    \\
\end{tabular}
\end{center}
\end{table*}


\renewcommand{\baselinestretch}{1.4}

\setcounter{qub}{0}

\begin{table*}[h]

\begin{center}
\caption{\label{Tab5} Coordinates, redshifts and magnitudes of QSOs}

\begin{tabular}{rlcclrl} \hline \\[-0.35cm]
\MC{1}{c}{\#}               &
\MC{1}{c}{Name}             &
\MC{1}{c}{$\alpha\,(1950)$} &
\MC{1}{c}{$\delta\,(1950)$} &
\MC{1}{c}{$z^{\,a}$}        &
\MC{1}{c}{$m_{B}$}          &
\MC{1}{c}{Detected emission lines }\\
&
\MC{1}{c}{ (2) } &
\MC{1}{c}{ (3) } &
\MC{1}{c}{ (4) } &
\MC{1}{c}{ (5) } &
\MC{1}{c}{ (6) } &
\MC{1}{c}{ (7) }
\\ \hline
\qq & HS 1138+4900  &11 38 13.4 &+49 00 09  &3.353 &18.9  & Ly$\alpha$ 1216 \AA, Si{\sc iv}/O{\sc iv}] 1400 \AA, C{\sc iv} 1549 \AA \\ 
\qq & HS 1203+3811  &12 03 53.7 &+38 11 09  &3.184 &$^+$19.6  & Ly$\alpha$ 1216 \AA                                                     \\ 
\qq & HS 1250+4408  &12 50 27.1 &+44 08 53  &1.639 &$^+$18.4 & C{\sc iv} 1549 \AA, C{\sc iii}] 1909 \AA, Mg{\sc ii} 2798 \AA           \\ 
\qq & HS 1334+3948  &13 34 52.3 &+39 48 15  &3.223 &19.1  & Ly$\alpha$ 1216 \AA, Si{\sc iv}/O{\sc iv}] 1400 \AA, C{\sc iv} 1549 \AA \\ 
\qq & HS 1528+3914  &15 28 39.4 &+39 14 25  &3.290 &17.0  & Ly$\alpha$ 1216 \AA, Si{\sc iv}/O{\sc iv}] 1400                           \\ 
\qq & HS 1608+3546  &16 08 00.5 &+35 46 30  &0.829 &17.8  & Mg{\sc ii} 2798 \AA                                                     \\ 
\qq & HS 1646+4501  &16 46 19.0 &+45 01 32  &1.790 &19.0  & C{\sc iv} 1549 \AA, C{\sc iii}] 1909 \AA                                \\ 
\qq & HS 1714+4202  &17 14 46.3 &+42 02 49  &0.846 &$^+$18.1 & Mg{\sc ii} 2798 \AA                                                   \\ 

\hline \\[-0.2cm]
\multicolumn{7}{l}{ $^a$ Observed redshift; $^+$ corrected APM magnitudes (see Sect. \ref{brief}).} \\
\end{tabular}
\end{center}
\end{table*}

\renewcommand{\baselinestretch}{1.4}

\setcounter{qub}{0}

\begin{table*}[h]

\begin{center}
\caption{\label{Tab6} Galaxies without detected emission lines }

\begin{tabular}{rlllrccl} \hline \\[-0.35cm]
\multicolumn{1}{c}{\#}   &
\multicolumn{1}{c}{Name} &
\multicolumn{1}{c}{$\alpha\,(1950)$} &
\multicolumn{1}{c}{$\delta\,(1950)$} &
\multicolumn{1}{c}{$v^{\,a}_{0}$} &
\multicolumn{1}{c}{m$_B$} &
\multicolumn{1}{r}{M$_B^{\,b}$} &
\multicolumn{1}{c}{Absorption lines} \\
&
\MC{1}{c}{ (2) } &
\MC{1}{c}{ (3) } &
\MC{1}{c}{ (4) } &
\MC{1}{c}{ (5) } &
\MC{1}{c}{ (6) } &
\MC{1}{c}{ (7) } &
\MC{1}{c}{ (8) }
\\ \hline
\qq& HS 0716+4510$^\dagger$  & 07 16 35.8 &+45 10 09  &33994 &19.2 &--19.0  & H$\alpha$, NaD                              \\
\qq& HS 0805+4120$^\dagger$  & 08 05 01.7 &+41 20 23  &17071 &$^+$17.8&--18.9  & H$\beta$, H$\gamma$, G$_{band}$, CaK, CaH   \\
\qq& HS 1010+4906$^{1}$  & 10 10 00.3 &+49 06 56  &18070 &19.3 &--17.6 &CaK, CaH, G$_{band}$, MgIb , NaD, probable emis. H$\alpha$   \\ 
\qq& HS 1052+3637$^\dagger$  & 10 52 32.9 &+36 37 05  & 7774 &19.0 &--16.1  & CaK, CaH                                    \\
\qq& HS 1350+4701$^\dagger$  & 13 50 52.3 &+47 01 41  &18570 &17.7&--19.2  & H$\alpha$                                   \\
\qq& HS 1420+4624$^\dagger$  & 14 20 21.1 &+46 24 39  & 9276 &18.5&--16.9  & H$\alpha$, H$\beta$, MgIb                   \\
\qq& HS 1443+4253  & 14 43 07.8 &+42 53 13  & 5082 &$^+$18.1&--16.1  & CaK, CaH, H$\beta$, MgIb, TiO 6155-6180     \\
\qq& HS 1454+3513  & 14 54 22.4 &+35 13 30  &35356 &$^+$19.1&--19.2  & H$\alpha$, H$\beta$, CaK, CaH, MgIb         \\
\qq& HS 1527+4252$^\dagger$  & 15 27 06.3 &+42 52 55  &35360 &$^+$17.5&--20.8  & H$\alpha$, H$\beta$, H$\gamma$, G$_{band}$, \\
\hline \\[-0.2cm]
\multicolumn{8}{l}{ $^a$ Heliocentric velocities; $^b$ absolute magnitudes are not corrected for galactic extinction;} \\
\multicolumn{8}{l}{ $^\dagger$ objects from the APM second priority sample; $^+$ APM magnitude.} \\
\multicolumn{8}{l}{ $^{1}$ a galaxy near HS 1009+4906 and HS 1010+4907 in Table 4.} \\

\end{tabular}
\end{center}

\end{table*}

\setcounter{qub}{0}

\begin{table*}[t]
\begin{center}
\caption{\label{Tab7}Stars}
\begin{tabular}{rlccrcl} \hline \\[-0.35cm]
\multicolumn{1}{c}{\#}             &
\multicolumn{1}{c}{Name}             &
\multicolumn{1}{c}{$\alpha\,(1950)$} &
\multicolumn{1}{c}{$\delta\,(1950)$} &
\multicolumn{1}{r}{m$_{B}$}          &
\multicolumn{1}{l}{Type}             &
\multicolumn{1}{c}{Spectral features} \\
&
\MC{1}{c}{ (2) } &
\MC{1}{c}{ (3) } &
\MC{1}{c}{ (4) } &
\MC{1}{c}{ (5) } &
\MC{1}{c}{ (6) } &
\MC{1}{c}{ (7) }
\\ \hline

\qq &HS 0712+4338$^\ddagger$   &07 12 46.4 &+43 38 30  &18.0   &F-G   & H$_{\alpha}$, H$_{\beta}$, H$_{\gamma}$, CaK, CaH                           \\
\qq &HS 0721+4148  &07 21 16.0 &+41 48 55  &18.0   &?     & H$_{\alpha}$, NaD, CaK, CaH                                                 \\
\qq &HS 0729+4559$^\ddagger$   &07 29 08.8 &+45 59 03  &18.0   &F-G   & H$_{\alpha}$, NaD,H$_{\beta}$, H$_{\gamma}$                                 \\
\qq &HS 0742+4205$^\ddagger$   &07 42 11.9 &+42 05 38  &17.9  &F     & H$_{\alpha}$, H$_{\beta}$, H$_{\gamma}$,CaK, CaH, NaD, MgIb                 \\
\qq &HS 1019+4002$^\ddagger$   &10 19 36.5 &+40 02 20  &17.2   &G     & CaK, CaH, G$_{band}$, H$_{\gamma}$, H$_{\beta}$, MgIb                       \\
\end{tabular}
\end{center}
\end{table*}

\clearpage



\end{document}